\documentclass[hidelinks,12pt]{article}
\usepackage{fullpage}
\usepackage[utf8]{inputenc}
\usepackage{amsmath, amssymb, url, float, graphicx, float}
\usepackage{cite}

\newcommand{\downto}{\downarrow}

\renewcommand\phi\varphi
\newcommand{\eps}{\varepsilon}

\let\phi\varphi

\newcommand{\ones}{\mathbf 1}
\newcommand{\reals}{{\mbox{\bf R}}}

\newcommand{\cf}{{\it cf.}}
\newcommand{\eg}{{\it e.g.}}
\newcommand{\ie}{{\it i.e.}}

\newcommand{\BEAS}{\begin{eqnarray*}}
\newcommand{\EEAS}{\end{eqnarray*}}
\newcommand{\BEA}{\begin{eqnarray}}
\newcommand{\EEA}{\end{eqnarray}}
\newcommand{\BEQ}{\begin{equation}}
\newcommand{\EEQ}{\end{equation}}
\newcommand{\BIT}{\begin{itemize}}
\newcommand{\EIT}{\end{itemize}}

\newif\iftodos

\todostrue

\title{Concave Pro-rata Games}
\author{Nicholas A.\ G.\ Johnson\\ 
    {\small\texttt{nagj@mit.edu}}
    \and Theo Diamandis\\
    {\small\texttt{tdiamand@mit.edu}}
    \and Alex Evans\\
    {\small\texttt{aevans@baincapital.com}}
    \and Henry de Valence\\
    {\small\texttt{hdevalence@penumbra.zone}}
    \and Guillermo Angeris\\
    {\small\texttt{gangeris@baincapital.com}}
}
\date{October 2022}

\begin{document} 
\maketitle 

\begin{abstract}
    In this paper, we introduce a family of games called concave pro-rata
    games. In such a game, players place their assets into a pool, and the pool
    pays out some concave function of all assets placed into it. Each
    player then receives a pro-rata share of the payout; \ie, each player receives
    an amount proportional to how much they placed in the pool. Such games
    appear in a number of practical scenarios, including as a simplified
    version of batched decentralized exchanges, such as those proposed by
    Penumbra. We show that this game has a number of interesting properties,
    including a symmetric pure equilibrium that is the unique equilibrium of
    this game, and we prove that its price of anarchy is $\Omega(n)$ in the number of
    players. We also show some numerical results in the iterated setting which
    suggest that players quickly converge to an equilibrium in iterated play.
\end{abstract}

\section*{Introduction}
Existing blockchain systems come to consensus on transactions in batches, called
blocks. Yet the economic mechanisms those transactions interact with are
generally designed to process each individual transaction sequentially, making
their behavior reliant on the ordering of transactions within the batch.  This
abstraction mismatch is the primary source of miner extractible value (MEV),
defined as economic value that can be captured by the block proposer (originally
the miner) who selects and sequences the transactions to be included in the
batch~\cite{flashboys2}.

%MEV, its impacts, and its potential mitigations are an active area of research.
%Most of that research has focused on ways to ...

However, rather than trying to blind the block proposer, or choose a ``fair''
ordering (which is difficult, if not impossible, to construct in any direct sense
on current systems)
%\footnote{The concept of a ``fair'' ordering for transactions the system
	%is coming to consensus on is necessarily incoherent: until the system has come
	%to consensus on the existence of the transactions, they do not exist, and once
	%it has, it is too late to define an ordering within the batch.}
within a batch, we could alternatively attempt to design economic mechanisms
which do not depend on the order of transactions within a block, and instead,
process each batch of transactions `all at once'. These mechanisms would then
be aligned with the actual ordering provided by the consensus mechanism,
stepping from one batch of transactions to the next in the same discrete time
steps in which consensus happens.

One such mechanism is a `pro-rata mechanism'. In this mechanism, there is some
known notion of value: for example, every user might want to trade some asset
$A$ for another, say $B$, and everyone `pitches in' some amount of asset $A$
into a pool. After everyone has placed their amounts, the pool, as a whole, is
traded on an exchange for some amount of asset $B$, and the resulting amount of
asset $B$ is distributed back to each player, in proportion to how much of
asset $A$ each player placed in the pool. It is not difficult to show that such
a mechanism has the desired property: the order in which players placed asset
$A$ into the collective pool does not change how much of asset $B$ each player
receives. Using some ideas from cryptography, this game can additionally be
implemented in a way that does not reveal any one player's contributions or
identity~\cite{zswap}, and so may be considered a simultaneous game.

On the other hand, mechanisms of this form often lead to interesting phenomena
as users must now consider the possible actions of other users when planning
their own actions. A natural framework to study these kinds of problems, where
players must reason about the strategies of other players, recursively, is via
game theory and the study of the equilibria of games~\cite{von2007theory}.
This paper serves to cleanly set up the game resulting from a pro-rata mechanism
in a simple mathematical framework and derive a number of useful results for such
games.

\paragraph{This paper.} The paper is organized as follows. We introduce the
concave pro-rata game in~\S\ref{sec:pro-rata-game} and show a few interesting
properties under mild conditions. Such properties include the existence and
uniqueness of a symmetric pure strategy equilibrium and an explicit way of
efficiently computing this equilibrium by solving a single variable, unimodal
optimization problem. We also show some simple bounds for the price of anarchy.
In~\S\ref{sec:dex} we then describe how this type of game connects to a
recent proposal for a batched decentralized exchange. We run a number of simulations in~\S\ref{app:numerics} and~\S\ref{app:additional_numerics}, illustrating the price of
anarchy and showing that in the iterated setting agents appear to converge
quickly to the specified equilibrium.

\section{The concave pro-rata game}\label{sec:pro-rata-game}
We will define the \emph{pro-rata game} with $n$ players as the game with the
following payoff for player $i=1, \dots, n$:
\begin{equation}\label{eq:payoff}
	U_i(x) = \frac{x_i}{\ones^Tx}f(\ones^Tx).
\end{equation}
Here, $f: \reals_+ \to \reals$ is some function satisfying $f(0) = 0$, while $x
\in \reals_+^n$ is a nonnegative vector whose $i$th entry is the action
performed by the $i$th player. We will say the game is a \emph{concave}
pro-rata game if the function $f$ is a concave function. This game has a simple
interpretation: every player `pitches in' some amount $x_i$ into a pool,
totaling $\ones^Tx$, and the pool pays out
$f(\ones^Tx)$ depending only on the total amount pitched in by all players. The amount
paid out by the pool is then distributed among the players in a pro-rata way;
\ie, each player $i$ receives an amount proportional to how much she put into
the pool. For the remainder of this paper, we will assume that the function $f$
is concave. We note that concave pro-rata games consist of a special case of aggregative games in which the payoff of each player is a function of their strategy and the sum of the strategies of all players (\cf,~\cite{aggregativeGames}).

\paragraph{Concavity.} The payoff $U_i$ is concave in the $i$th entry, holding
the remaining entries constant. To see this, first define $y = \ones^Tx - x_i$
(\ie, $y$ is the sum of all entries of $x$ except the $i$th entry). Overloading
notation slightly for $U_i$, we have that
\[
U_i(x_i, y) = \frac{x_i}{x_i + y}f(x_i + y).
\]
We can write $U_i(\cdot, y)$ as the composition of the following two functions
\[
U_i(x_i, y) = g(x_i, h(x_i)),
\]
where
\[
g(x_i, t) = t f\left(\frac{x_i}{t}\right) \qquad \text{and} \qquad h(x_i) = \frac{x_i}{x_i+y},
\]
which are defined for nonnegative real inputs.
We will use this rewriting to show that this function is concave in $x_i$, since,
using the basic convex composition rules (\cf,~\cite[\S3.2.4]{cvxbook}) it
suffices to show that (a) $h$ is concave and (b) $g$ is concave and
nondecreasing in its second argument.

First, note that $h$ is (strictly) concave since
\[
h(x_i) = 1 - \frac{y}{x_i + y},
\]
which is evidently (strictly) concave in $x_i$
since $y$ is a constant. We can see that $g$ is jointly concave in its arguments 
as it is the perspective transform of the function $f$, which preserves 
concavity (\cf,~\cite[\S3.2.6]{cvxbook}). Finally, we need to show that $g$ is 
nondecreasing in its second argument. To see this, let $0 \le t \le t'$, then
we have
\begin{equation}\label{eq:proof}
	\begin{aligned}
		g(x_i, t') = t'f\left(\frac{x_i}{t'}\right) &= t'f\left(\frac{t}{t'}\frac{x_i}{t} + \left(1-\frac{t}{t'}\right)0\right)\\
		&\ge tf\left(\frac{x_i}{t}\right) + \left(1-\frac{t}{t'}\right)f(0) = tf\left(\frac{x_i}{t}\right) = g(x_i, t).
	\end{aligned}
\end{equation}
The inequality follows from the definition of concavity, while the second-to-last equality follows from the fact that $f(0) = 0$.

\paragraph{Selfish maximum.} The fact that $g$ is nondecreasing in its second
argument also has an interesting consequence: a player never does better in the
pro-rata game when compared to the `selfish' version.
In other words, for a fixed $x_i$, player $i$ has the largest payoff when all
other players $j \ne i$ have $x_j = 0$. This is easy to see since
\[
t = \frac{x_i}{\ones^Tx} \le 1
\]
so
\[
U_i(x) = g(x_i, t) \le g(x_i, 1) = f(x_i),
\]
where $g$ is as defined above. The inequality follows from the monotonicity of
$g$ in its second argument.

\paragraph{Strict concavity.} In the important special case where $f$ satisfies
\begin{equation}\label{eq:strict}
	f(\alpha t) > \alpha f(t),
\end{equation}
for every $t > 0$ and $0 < \alpha < 1$, then the function $U_i(\cdot, y)$ is strictly
concave in its first argument. (We will show this soon.)
Property~\eqref{eq:strict} has the interpretation that any chord of the
function, drawn between $(0,0)$ and any other point on the graph, lies strictly
below the function itself. For example, a sufficient condition is that the
function $f$ is strictly concave, though this condition is not a necessary one
as there are functions which are not strictly concave that
satisfy~\eqref{eq:strict}. See appendix~\ref{app:strict-concavity} for a more general
condition.

Since we know that $h$ is a strictly concave function and $g$ is a concave
function, we can show that $g(x_i, h(x_i))$ is strictly concave in $x_i$ by 
showing that $g$ is strictly increasing in its second argument.
Strict concavity of $g$ follows from the usual composition rules 
(see~\cite[\S3.2.4]{cvxbook}). 
To show that $g$ is strictly increasing in its second
argument, let $0 < t < t'$, then:
\[
g(x_i, t') = t'f\left(\frac{x_i}{t'}\right) =
t'f\left(\frac{t}{t'}\frac{x_i}{t}\right) >
t'\frac{t}{t'}f\left(\frac{x_i}{t}\right) = tf\left(\frac{x_i}{t}\right) =
g(x_i, t),
\]
where the inequality follows from an application of~\eqref{eq:strict} with
$\alpha = t/t'$.

\paragraph{Definitions.} For completeness, we state several important game
theoretic definitions~\cite{von2007theory}. To each player $i=1, \dots, n$, we associate a
\emph{strategy}, which is a probability distribution $\pi_i$ over the possible
actions of player $i$, the nonnegative real numbers. We say a strategy is
\emph{pure} if $\pi_i$ is a deterministic distribution or point mass. In other
words, we say a strategy is pure when the probability of choosing a specific
action $z$ is always one; \ie, $\pi_i(\{z\}) = 1$ for some $z\ge 0$. Otherwise,
we say the strategy is \emph{mixed}.

A \emph{Nash equilibrium} (simply an \emph{equilibrium} from here on out) is a
collection of strategies $\pi_i$ for each player $i=1, \dots, n$ such that no
individual player can achieve a strictly better outcome by choosing a different
strategy. Concretely, let $x_i \sim \pi_i$ be a random variable chosen by
player $i$'s strategy (mixed or pure) and let $y_i \sim \pi_{-i}$ be a random
variable denoting the sums of random variables from other players' strategies.
The collection of strategies $(\pi_i)$ consists of an equilibrium if, for each
player $i$, we have
\[
\mathbf{E}_{x_i \sim \pi_i, ~y_i \sim \pi_{-i}}\left[{U_i(x_i,
	y_i)}\right] \geq \mathbf{E}_{x_i \sim \tilde\pi_i, ~y_i \sim
	\pi_{-i}}\left[{U_i(x_i, y_i)}\right],
\]
where $\tilde \pi_i$ denotes any strategy. (For the remainder of the paper, we
will drop the $x_i$ and $y_i$ in the definition of the expectation to shorten
notation.) If the above condition holds with strict inequality for all $i$
except when $\tilde \pi_i = \pi_i$, the equilibrium is said
to be \emph{strict}. In words, an equilibrium is strict if each player would
achieve a strictly worse outcome by choosing a different strategy. In
general, we say an equilibrium is pure if all strategies of that equilibrium
are pure, and mixed otherwise.

\paragraph{Pure equilibria.} With those definitions, we note that the strict
concavity of $U_i(x_i, y)$ in $x_i$ has an important, direct consequence: every
equilibrium of this game is a pure equilibrium. Let $x_i \sim \pi_i$ be any
strategy that is not pure, while $y_i \sim \pi_{-i}$ is a random variable
denoting the sums of the other players' strategies, then
\[
\mathbf{E}_{\pi_i, \pi_{-i}}\left[{U_i(x_i, y_i)}\right] = 
\mathbf{E}_{\pi_{-i}} \left[ \mathbf{E}_{\pi_{i}} \left[{U_i(x_i, y_i)} \right] \right] <
\mathbf{E}_{\pi_{-i}} \left[ {U_i(\mathbf{E}_{\pi_{i}}[x_i], y_i)} \right],
\]
where the strict inequality is a result of strict concavity of $U_i(\cdot, y)$
for all $y$ and the fact that $\pi_i$ is not a point mass. In other words, if
$x_i \sim \pi_i$ is a mixed strategy for player $i$, then this player is always
strictly better off playing the pure strategy $\mathbf{E}_{\pi_i}[x_i]$
instead. For the remainder of this paper, we will assume that $f$ is concave
and satisfies condition~\eqref{eq:strict}, unless otherwise stated.
Additionally we will only discuss pure equilibria for the remainder of the
paper, as all equilibria must be pure, so talking about a strategy as
as a specific action $x_i \in \reals_+$ is reasonable.

\paragraph{Extensions.} A simple immediate extension to the concave pro-rata
game is to consider payoff functions of the form:
\[
U_i(x) = \frac{c_ix_i}{c^Tx}f\left(c^Tx\right),
\]
for some strictly positive vector $c \in \reals_{++}^n$. A more general
extension is when we have a collection of $n$ strictly increasing functions
$\phi_i: \reals_+ \to \reals$, where $\phi_i(0) = 0$ and $\phi_i(t) \to \infty$
when $t \to \infty$ for $i=1, \dots, n$, and
\[
U_i(x) = \left(\frac{\phi_i(x_i)}{\sum_{i=1}^n \phi_i(x_i)}\right)f\left(\sum_{i=1}^n \phi_i(x_i)\right).
\]
In either case, all of the same properties given above apply to this slightly
more general game with nearly identical proofs, but we will only consider the
(often useful) special case where $\phi_i(t) = t$.

\subsection{Symmetric pure strict equilibrium} \label{sec:symmetric_equi}
There is a strict, pure equilibrium where all players have equal strategies,
given by $x = (q/n) \ones$ where $q$ is the optimizer of the following problem:
\begin{equation}\label{eq:shared-problem}
	\begin{aligned}
		& \text{maximize} && q^{n-1}f(q)\\
		& \text{subject to} && q \ge 0,
	\end{aligned}
\end{equation}
with variable $q \in \reals$. We will show some properties of this result
first and then show that the pure strategy $x = (q/n)\ones$ is, indeed,
an equilibrium.

\paragraph{Solution properties.} This problem has a (unique) and finite
solution $q > 0$ provided $f(z) > 0$ and $f(w) = 0$ for some $0 < z < w$. 
%In the first case, the problem becomes trivial and there is no
%finite equilibrium since every agent $i$ could increase their strategy and
%receive a larger payoff, in the latter, any choice of strategy is an
%equilibrium as they all have $0$ payoff.
The fact that $q > 0$ follows by
noting that that $q$ must satisfy $q^{n-1} f(q) \ge z^{n-1}f(z) > 0$ since it
is optimal. On the other hand, the fact that $q$ is finite follows from the
fact that, for any $r \ge w$, we have
\[
\frac{w}{r}f(r) + \left(1 - \frac{w}{r}\right)f(0) \le f(w) = 0,
\]
where the first inequality follows from the definition of concavity. Since, by
assumption, $f(0) = 0$ and $w/r > 0$, we have that $f(r) \le 0$ so $r$ cannot
be optimal. (In fact, both statements combined prove the stronger fact that $0
< q < w$, but this is not necessary for what follows.) The uniqueness of the
solution to problem~\eqref{eq:shared-problem} follows from observing that the
logarithm of the objective function is strictly concave. (This is true since
$\log$ is strictly increasing and $\log\circ f$ is concave if $f$ is concave.)

\paragraph{Discussion.} It may appear that the condition placed on $f$ is very
strong, but in fact, any $f$ not satisfying the above condition has only
trivial (or no) equilibria. In particular, since $f$ is concave, if $f$ does
not satisfy the above condition, either (a) $f$ is strictly positive everywhere
except at $f(0)=0$, (b) $f$ is strictly negative everywhere except at $f(0)=0$,
or (c) $f = 0$. In the first case, there is no equilibrium as any player can
improve their payoff by increasing their strategy. In the second case, any
player who plays a nonzero strategy receives negative payoff (whereas playing
the zero strategy would give 0 payoff). While, in the third case, any strategy
is an equilibrium.

\paragraph{Equilibrium properties.} The collection of strategies $x =
(q/n)\ones$ is clearly pure and symmetric. To see that $x = (q/n)\ones$ is a
strict equilibrium, note that the best response for any player $i$, when every
other player plays strategy $q/n$ is:
\begin{equation}\label{eq:best-response}
	\begin{aligned}
		& \text{maximize} && \frac{x_i}{x_i + (1-1/n)q}f(x_i + (1-1/n)q)\\
		& \text{subject to} && x_i \ge 0,
	\end{aligned}
\end{equation}
with variable $x_i \in \reals$. We will show that the solution to 
\eqref{eq:best-response} is $x_i = q/n$ in two steps. First, we will
show that any solution must have $x_i > 0$ and therefore that the first order
optimality conditions applied to the objective suffice. We will then show that
$x_i = q/n$ is a solution to the optimality conditions. This result, combined with the
fact that the objective is strictly concave, implies that $x_i = q/n$ is the
unique solution to the optimality conditions, which proves the final claim that
this equilibrium is strict.

To see that any solution to the best response problem~\eqref{eq:best-response}
must have $x_i > 0$, note that $q/n$ is feasible and achieves an objective
value of $f(q)/n > 0$, which is strictly greater than the objective value of
zero achieved by $x_i = 0$.

Next, note that $q > 0$ must satisfy the first order 
optimality conditions of~\eqref{eq:shared-problem}:
\begin{equation}\label{eq:opt-condition}
	(n-1)f(q) + qf'(q) = 0.
\end{equation}
On the other hand, the first order optimality conditions for the objective of
problem~\eqref{eq:best-response} are that $x_i$ must satisfy (writing $q' = (1
- 1/n)q$ for convenience)
\[
\frac{q'}{x_i + q'}f(x_i + q') + x_if'(x_i + q') = 0.
\]
Choosing $x_i = q/n$ clearly satsfies this condition, since plugging
this value in gives
\[
\left(1 - \frac1n\right)f(q) + \frac{qf'(q)}{n} = \frac1n((n-1)f(q) + qf'(q)) = 0,
\]
as required. Since the objective is strictly concave, this is the unique $x_i$
satisfying the optimality conditions and is therefore the best response.
Additionally, while we have assumed that $f$ is differentiable, a very similar
proof using subgradient calculus gives an identical result.

\subsection{Uniqueness of equilibrium}\label{sec:unique}
In fact, it is not hard to show that the symmetric, pure, strict equilibrium
is, surprisingly, the unique equilibrium for this game, under the same
conditions as~\eqref{eq:shared-problem}; \ie, that $f(z) > 0$ and $f(w) = 0$
for some $0 < z < w$. This proof can be broken down into a few steps. First, we
will show that any equilibrium $x$ satisfies $f(\ones^Tx) > 0$ and $x_i > 0$
for each $i$. This will then be used to show that there is no non-symmetric
equilibrium, and, since we know that any symmetric equilibrium must satisfy
equation~\eqref{eq:shared-problem}, which has a unique solution, we then know
that it is the unique equilibrium of this game.

\paragraph{Positivity of equilibria.} First we will show that $f(v) > 0$ for
every $0 < v < w$. To see this, note that the function $f$ is bounded from
below by all of its chords, as it is a concave function. Note that the chord
with endpoints $(0, 0)$ and $(z, f(z))$ lies above the $x$-axis, except at
$(0,0)$, while the chord with endpoints $(z, f(z))$ and $(w, f(w)) = (w, 0)$
lies above the $x$-axis, except at $(w, 0)$, which leads to the final result.

Now, suppose a collection of (pure) strategies satisfies $f(\ones^Tx) < 0$.
Since $f(0) = 0$, there is some index $i$ such that $x_i > 0$. This implies
that $U_i(x_i, \ones^Tx - x_i) < 0$. But then player $i$ can achieve a payoff
equal to $0$ by employing the strategy $\tilde x_i = 0$, which is strictly
better than a negative payoff, so $x$ cannot be an equilibrium.

On the other hand, if a collection of strategies satisfies $f(\ones^Tx) = 0$,
then, from the previous discussion, we must have either $\ones^Tx = 0$ or
$\ones^Tx = w$. If $\ones^Tx = 0$, any player $i$ can obtain a strictly
positive payoff by playing the strategy $\tilde x_i = z$. If, instead,
$\ones^Tx = w > 0$, there is some index $i$ such that $x_i > 0$. We have that
the player's payoff is $U_i(x_i, \ones^Tx - x_i) = 0$ which means that
\[
U_i(x_i - \eps, \ones^Tx - x_i) = \frac{x_i - \eps}{\ones^Tx - \eps}f(\ones^Tx - \eps) > 0,
\]
for 
$\eps > 0$ small enough since $f(w - \eps) > 0$, so $x$ is not an
equilibrium. 

Putting all of these statements together means that any equilibrium $x$
satisfies $f(\ones^Tx) > 0$ and $\ones^Tx < w$. To see that any equilibrium
must also satisfy $x > 0$, note that if there exists an index $i$ with $x_i =
0$ for a collection of strategies with $f(\ones^Tx) > 0$, player $i$ can always
achieve a strictly positive payoff by playing $\tilde x_i = \eps > 0$,
for $\eps$ small enough. 

\paragraph{Symmetry of equilibria.}
Next, we will show that if $x_i$ is a best response for player $i$, then any
$j$ for which $x_j > x_i$ is not a best response for player $j$, and vice
versa. This will immediately show that any equilibrium must satisfy $x_i = x_j$
(\ie, it is symmetric). We will show this in the case that $f$ is
differentiable, but a similar proof holds in the more general case, using
subgradient calculus.

Let $x$ be an equilibrium with $x_j > x_i$. Given that $x_i$ is a best response, then the
optimality conditions for~\eqref{eq:best-response} imply that:
\[
\left(\frac{\ones^Tx - x_i}{(\ones^Tx)^2}\right)f(\ones^Tx) + \frac{x_i}{\ones^Tx}f'(\ones^Tx) = 0.
\]
Since $x$ is an equilibrium, from the previous discussion, we have that
$f(\ones^Tx) > 0$, $x_i > 0$, and $\ones^Tx > x_i$, so $f'(\ones^Tx) < 0$. On
the other hand, differentiating the objective of the best response
problem~\eqref{eq:best-response} for player $j$ gives
\[
\left(\frac{\ones^Tx - x_j}{(\ones^Tx)^2}\right)f(\ones^Tx) + \frac{x_j}{\ones^Tx}f'(\ones^Tx) < \left(\frac{\ones^Tx - x_i}{(\ones^Tx)^2}\right)f(\ones^Tx) + \frac{x_i}{\ones^Tx}f'(\ones^Tx) = 0,
\]
where the inequality follows from the fact that, since $x_i < x_j$ we have
\[
\frac{\ones^Tx - x_j}{(\ones^Tx)^2} < \frac{\ones^Tx - x_i}{(\ones^Tx)^2} \quad \text{and} \quad \frac{x_j}{\ones^Tx} > \frac{x_i}{\ones^Tx},
\]
so $x_j$ cannot be a best response as it is not optimal
for~\eqref{eq:best-response}. The converse case, when $x_j < x_i$ with $x_i$
being a best response, follows from a nearly identical proof. This immediately
implies that any equilibrium must be symmetric, so, from the preceding
discussion, the unique equilibrium is the one given by the solution to
problem~\eqref{eq:shared-problem}.

\subsection{Equilibrium payoff}
Conditioned on each player receiving the same payoff (a fairness
condition), the optimal allocation every player would get is
\[
\frac1n\sup f,
\]
which is, by definition, at least as good as the equilibrium payoff:
\[
\frac1n f(q),
\]
where $q > 0$ is the solution to~\eqref{eq:shared-problem}. In fact, we can
show that the optimal fair allocation is always strictly better than the
equilibrium payoff. To see this, note that, under the assumptions on $f$
introduced above, we know $\sup f$ is achieved by some value $0 < q^\star
< w$, satisfying $f'(q^\star) = 0$. Rearranging the first order optimality
condition for $q$ in problem~\eqref{eq:shared-problem} gives
\[
f'(q) = -(n-1)\frac{f(q)}{q} < 0,
\]
for all $n > 1$ since $f(q) > 0$. This means that $q$ does not satisfy the
optimality condition for maximizing $f$, so $f(q) < f(q^\star) = \sup f$. (In
fact, this says slightly more: using the concavity of $f$, we have that $q >
q^\star$, \ie, that players `overpay' at equilibria when $n > 1$.)

\paragraph{Price of anarchy.}
Given the same assumptions as the beginning of~\S\ref{sec:unique} on the
function $f$, it is not difficult to show that the price of anarchy satisfies
\[
\frac{\sup f}{f(q)} \geq \Omega(n)
\]
as the number of players $n$ becomes large for some constant $C$. To see this, consider the first
order optimality conditions for $y$~\eqref{eq:shared-problem}:
\[
(n-1)f(q) + qf'(q) = 0.
\]
Note that $f'(q) < 0$ since $q > 0$ and $f(q) > 0$, so
\[
f(q) = -\frac{qf'(q)}{n-1} > 0,
\]
whenever $n > 1$. Since $f$ is concave, then $f'$ is monotonically
nonincreasing, and, since $q \le w$ for
every $n$ we have that
\[
f(q) = -\frac{qf'(q)}{n-1} \le -\frac{wf'(w)}{n-1} \le O\left(\frac1n\right).
\]
Finally, we know that $\sup f$ is constant in the number of players, so
\[
\frac{\sup f}{f(q)} \geq \Omega(n).
\]

%\subsection{No-arbitrage condition}
%XXX: Write about how, if $f(z) = g(z) - cz$ for some $c > 0$ then how far
%is $g'(y)$ from $c$?
%
%Suppose the function $f$ is given by $f(z) = g(z) - cz$ where $c > 0$. The function $g(z)$ can be thought of as the value of the asset output by the CFMM while $c$ is the reference price of the asset in some external market. The required conditions on $f$ imply that $g(0) = 0$, $g$ is concave and that there exists $0 < q < w$ such that $g(q) > cq$ and $g(w) = cw$. 
%
%Suppose $\sup f$ is achieved by $\hat{y}$. We must have $f'(\hat{y}) = 0$  which implies that $g'(\hat{y}) = c$. In words, the marginal price of the asset in the CFMM is equal to the reference price under the optimal fair allocation. This is the no-arbitrage condition that arises in "traditional" CFMMs. This condition also holds for $n=1$, when a single player plays the concave pro-rate game. However, as $n$ grows, $g'(y)$ departs from $c$ where $y$ solves \eqref{eq:shared-problem}.
%
%In particular, we can show that $g'(y) < c$ for $n > 1$. Since $y$ solves \eqref{eq:shared-problem}, we have
%\[
%f'(y) = -(n-1)\frac{f(y)}{y} < 0,
%\]
%since $y > 0$ and $f(y) > 0$. Since $f'(y) = g'(y) - c$, this immediately implies that $g'(y) < c$. 
%
%We wish to bound the ratio $\frac{g'(y)}{c}$. Since $y$ must satisfy the optimality conditions for \eqref{eq:shared-problem}, we have
%\[
%g'(y) = -(n-1) \frac{g(y) - cy}{y} + c
%\]
%This implies that 
%\[
%\frac{g'(y)}{c} = 1 - (n-1) \bigg{(}\frac{g(y)}{yc}-1\bigg{)}
%\]
%
%\noindent
%XXX: Derive something interesting here (want explicit bound)

\section{Batched decentralized exchanges}\label{sec:dex}
In this section, we will show some basic applications of the above properties
to a batched decentralized exchange, which we describe below.

\paragraph{Decentralized exchanges.} A \emph{decentralized exchange} (or DEX,
for short) is a type of exchange that exists on a blockchain. Such exchanges
enable any agent to trade currencies without the need for a centralized
intermediary. In many cases, these exchanges are organized as constant function
market makers (see, \eg,~\cite{angeris2022} for a general introduction to this type of
exchange), a special type of automated market maker that uses a specific
function to price assets.

\paragraph{Batched DEXs.} A \emph{batched} decentralized exchange is a DEX
where the trades are batched before they are executed. Specifically, the trades are
aggregated in some way (depending on the type of batching performed) and then
traded `all together' through the DEX, before being disaggregated and passed
back to the users. Though the idea of a batched exchange has been proposed many
times in different contexts (see,
\eg,~\cite{budish2015} and~\cite{walther2018}),
presently, almost all major decentralized exchanges are not batched. Recent
work has suggested that batching is useful for
privacy~\cite{chitraDifferentialPrivacyConstant2022} and Penumbra~\cite{zswap}
has proposed a design for a fully-private decentralized exchange which makes
use of batching as a method for avoiding certain information
leakage~\cite{angerisNotePrivacyConstant2021}. We describe a very simplified
version of this proposal below, which will suffice for our discussion.

\paragraph{Batching design.} In this scenario, we have traders $i=1, \dots, n$
who all wish to trade some amount, say $\Delta_i \in \reals$ of asset $A$
for some other asset, which we will call asset $B$. In this case, negative
values of $\Delta_i$ denote that trader $i$ wishes to receive some
amount of asset $A$ (and will instead tender asset $B$ to the protocol). For
convenience, we will assume that $\ones^T\Delta > 0$, \ie, on net, traders want
more of asset $B$ than asset $A$. The batching protocol of penumbra first
clears all trades to get a nonnegative vector of `residual' trades $\Delta' \in
\reals_+^n$ with $\ones^T\Delta =\ones^T\Delta'$. (In other words, the protocol
does not generate debts in any one side.) We can view $\Delta'$ as the `excess
demand' for asset $B$ over $A$ and leave the mechanism for constructing
$\Delta'$ otherwise unspecified, requiring only the additional condition that,
if $\Delta \ge 0$, then $\Delta' = \Delta$. (This condition can be roughly
stated as: if the only trades are due to excess demand, then no clearing
happens.) The protocol then pools the residual demand, $\ones^T\Delta'$ and
trades it against a constant function market maker, represented by some
function $g$, to receive $g(\ones^T\Delta')$ of asset $B$, 
which it then distributes to each
agent $i$ in a pro-rata way, leading to an identical form to that of the
pro-rata payoff~\eqref{eq:payoff} with $x = \Delta'$.
Constant function market makers always have concave $g$, known sometimes
as the \emph{forward exchange function} (\cf,~\cite[\S4]{angeris2022}), with
$g(0) = 0$, and, in many practical cases, these functions are strictly concave.

\subsection{Arbitrage}
A common way of analyzing markets is through the lens of \emph{arbitrage}: the
ability to exploit price differences in order to make essentially risk-free
profit. From before, we will write $g$ for the forward exchange function of a
constant function market maker, used by the batching design presented above.

\paragraph{Existence.} Assuming $g$ is differentiable at $0$, we
can interpret $g'(0)$ as the marginal amount of asset $B$ that one would
receive for a marginal amount of $A$. (The function $g$ is often not
differentiable at $0$, but is one-sided differentiable at $0^+$, which
suffices.) If $g'(0)$ is larger than the price of an external market, say $c >
0$, then anyone who can directly trade with $g$ can make risk-free profit by
trading some (potentially small) amount, $t > 0$ of asset $A$ for $g(t)$ of
asset $B$, and then sell this amount of asset $B$ to get $g(t)/c - t > 0$ of
profit. (One simple way to see this is true is to use the definition of a
derivative on $g(t)/c$ and send $t \downto 0$.)

%To see the fact that the profit is strictly positive for $t$ small enough, note
%that
%\[
%    f(t) \approx f(0) + f'(0)t + o(t) = f'(0)t + o(t).
%\]
%as $t$ becomes small, so,
%\[
%    \frac{f(t)}{c} \approx \frac{f'(0)}{c}t + o(t) > t + o(t).
%\]
%where the inequality follows from the assumption that $f'(0) > c$. Making $t$
%small enough means that the right-hand $o(t)$ term vanishes more quickly than
%any other term in the expression, so we are left with $f(t)/c > t$ for $t$
%small enough.

\paragraph{Optimal arbitrage.} Since an agent can make risk-free
profit in these cases, it is reasonable to ask: what is the maximum amount of
profit an agent can make with this strategy? This is known as the
\emph{optimal arbitrage problem}, written:
\[
\begin{aligned}
	& \text{maximize} && g(t) - ct\\
	& \text{subject to} && t \ge 0,
\end{aligned}
\]
with variable $t \in \reals$. From before, if we know that $g'(0) > c$, then
this problem has an optimal value that is strictly positive. If $g$ is
differentiable, the optimal solution $t^\star$ satisfies
\[
g'(t^\star) = c,
\]
which we can see from the first-order optimality conditions for this problem.
This has the interpretation that the marginal price of the CFMM after the trade
$t^\star$, given by $g'(t^\star)$ should be equal to the price of the external
market, which we defined to be $c$.

\paragraph{The (aggregated) arbitrage game.} In the batched exchange above,
arbitrageurs cannot directly trade with the constant function market maker, but
must instead go through the batching process. Assuming there are $n$ arbitrageurs
competing to maximize their profit, the next question is: what are the
properties of this game? Defining
\[
f(t) = g(t) - ct,
\]
then this game is a concave pro-rata game with function $f$, since the
payoff~\eqref{eq:payoff} for player $i$ is
\[
U_i(x_i, y_i) = \frac{x_i}{x_i + y_i}f(x_i + y_i) = \frac{x_i}{x_i + y_i}g(x_i + y_i) - cx_i.
\]
Note that this is exactly the amount received from the DEX with forward
exchange function $g$, minus the cost of trading $x_i$ with the external
market, for player $i$. This game inherits all of the properties derived
in~\S\ref{sec:pro-rata-game}. We show some numerical simulations of iterated
behavior for some utility functions of this form in
appendices~\ref{app:numerics} and~\ref{app:additional_numerics}.

\section{Conclusion}
We introduced concave pro-rata games and established several useful properties
under relatively mild conditions. In particular, we showed the existence of a
unique equilibrium that is symmetric and pure. This equilibrium can be computed
efficiently by solving a single variable, unimodal optimization problem. We
further established that the price of anarchy is $\Omega(n)$ in the number of
players, relative to the optimal `fair' allocation. We illustrated how concave
pro-rata games connect to a recent proposal for a batched decentralized
exchange and numerically studied the behavior of agents engaged in such a game
in the iterated setting for a specific form of utility function. Future work
includes further study of the optimal arbitrage problem for batched
decentralized exchanges.

\bibliographystyle{alpha}
\bibliography{citations.bib}

\appendix
\section{Numerics} \label{app:numerics}

The results of~\S\ref{sec:pro-rata-game} provide insight into the equilibrium
behavior of concave pro-rata games. Here we explore the transient behavior of
such games through simulation.

\paragraph{Game setup.} Suppose that the game is played
iteratively, and, at each iteration $t$, player $i$ chooses some action $x_i^t$
as the best response to the actions chosen by the other players in the previous
round (denoted as $x_{-i}^{t-1}$), possibly subject to additional constraints.
We consider the following scenarios:
\begin{enumerate}
	\item At iteration $t$, player $i$ takes action equal to the best response to $x_{-i}^{t-1}$.
	\item At iteration $t$, player $i$ takes action equal to the best response to $x_{-i}^{t-1}$
	subject to a budget constraint ($x_i^t \in [0, M_i]$).
\end{enumerate}

\paragraph{Payoff functions.} For these simulations, we use functions $f$ of
the form $f(t) = g(t)- c t$ where $c > 0$ and $g(t) = \frac{\gamma R_2 t}{R_1+\gamma t}$
with $0 < \gamma \leq 1$, $R_1, R_2 > 0$. The function $g(t)$ is the forward exchange function
for a Uniswap V2 swap pool with reserves $R \in \reals^2_+$ and fee parameter
$\gamma$ when asset $1$ is being tendered and asset $2$ is being received. This setting
simulates $n$ arbitrageurs competing to maximize their profit, where $c$ denotes the
external market price of asset $2$. For simulations using a somewhat more simple payoff function,
see appendix \ref{app:additional_numerics}. Note that $f$ is strictly concave
and therefore satisfies condition~\eqref{eq:strict}, and clearly $f(0) = 0$. 

\paragraph{Shared equilibrium.} 
The (unique) symmetric pure equilibrium strategy is the solution to
problem~\eqref{eq:shared-problem}. This is easy to compute using the first
order optimality conditions for problem~\eqref{eq:shared-problem} given
in~\eqref{eq:opt-condition}. Plugging in this particular form of $f$, we obtain
the following quadratic equation:
\begin{equation}\label{eq:best-response-quad}
	(cn\gamma^2)q^2 + q(\gamma^2 R_2 + 2cnR_1\gamma-\gamma^2nR_2)+(cnR_1^2-\gamma nR_1R_2) = 0.
\end{equation}
The equilibrium is then given by $x_i = q/n$, for each player $i=1, \dots, n$ where
$q$ denotes the positive root of~\eqref{eq:best-response-quad}.

\paragraph{Best responses.} The best response of
player $i$, given the budget constraint $0 \leq x_i \leq M_i$ and other players'
strategies $y_i = \ones^Tx - x_i$, is given by
\begin{equation}\label{eq:best-response-sim}
	\begin{aligned}
		& \text{maximize} && U(x_i, y_i)\\
		& \text{subject to} && x_i \in [0, M_i],
	\end{aligned}
\end{equation}
with variable $x_i \in \reals$. This is a single-variable convex optimization
problem that is easily solved in practice by any number of off-the-shelf
packages~\cite{diamond2016cvxpy, ocpb:16}. When $x_i$ is unconstrained,
the optimal value of~\eqref{eq:best-response-sim} is given by
\[
x_i = \frac{1}{\gamma}\left(\sqrt{\frac{\gamma R_1 R_2 + \gamma^2 R_2 y}{c}} -R_1\right)- y
\]
For more details, the code is available at (anonymized for review).
%\begin{center}
%    \texttt{github.com/bcc-research/penumbra-arb/experiments/transient\_sims.ipynb}
%\end{center}

\begin{figure}
	\centering
	\includegraphics[width=0.8\linewidth]{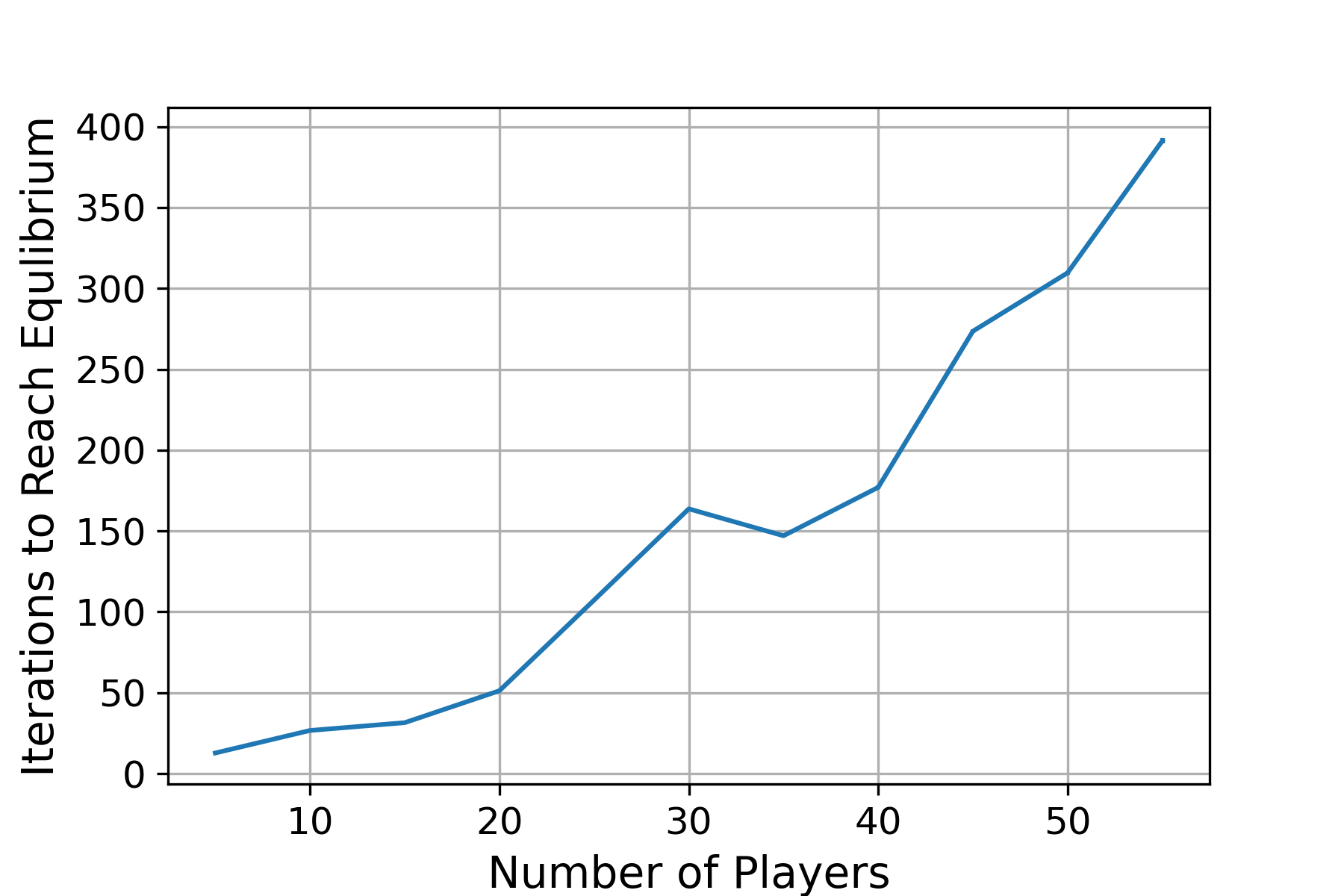}
	\caption{Number of iterations to reach equilibrium versus the number
		of players in Scenario 1.}
	\label{fig:scenario1+2}
\end{figure}
\paragraph{Simulation results.} In our simulations, we fix $\gamma=0.99$, $R_1 = 200$,
$R_2 = 250$, and $c=1$. We average each reported value over $100$ trials. In
figure~\ref{fig:scenario1+2}, the intial strategy of each player is drawn
uniformly at random from the interval $(0, w/n)$, where $w$ is a value such
that $f(w) = 0$.

Figure~\ref{fig:scenario1+2} illustrates that the number of
iterations needed to reach the unique equilibrium, in the absence of budget
constraints, scales superlinearly in the number of players. We define the number
of iterations to reach equilibrium as the number iterations until the strategy of
every player is equal to the unique equilibrium up to the first decimal place;
\ie, the first round $t$ such that 
\[
\max_i |x_i^t - x^\star| < 0.1,
\]
where $x^\star$ denotes the equilibrium strategy.

\begin{figure}
	\centering
	\includegraphics[width=0.45\linewidth]{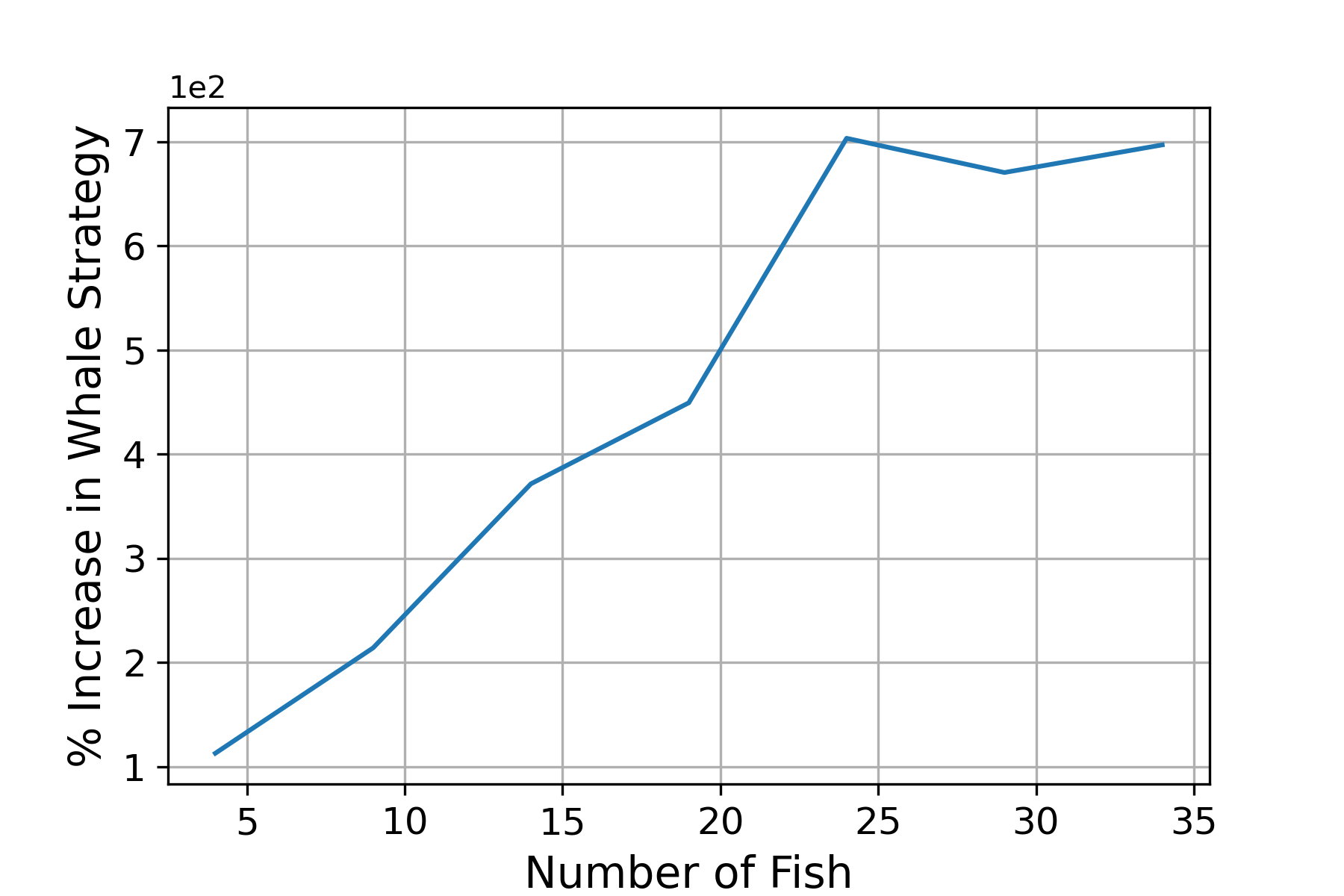}
	\includegraphics[width=0.45\linewidth]{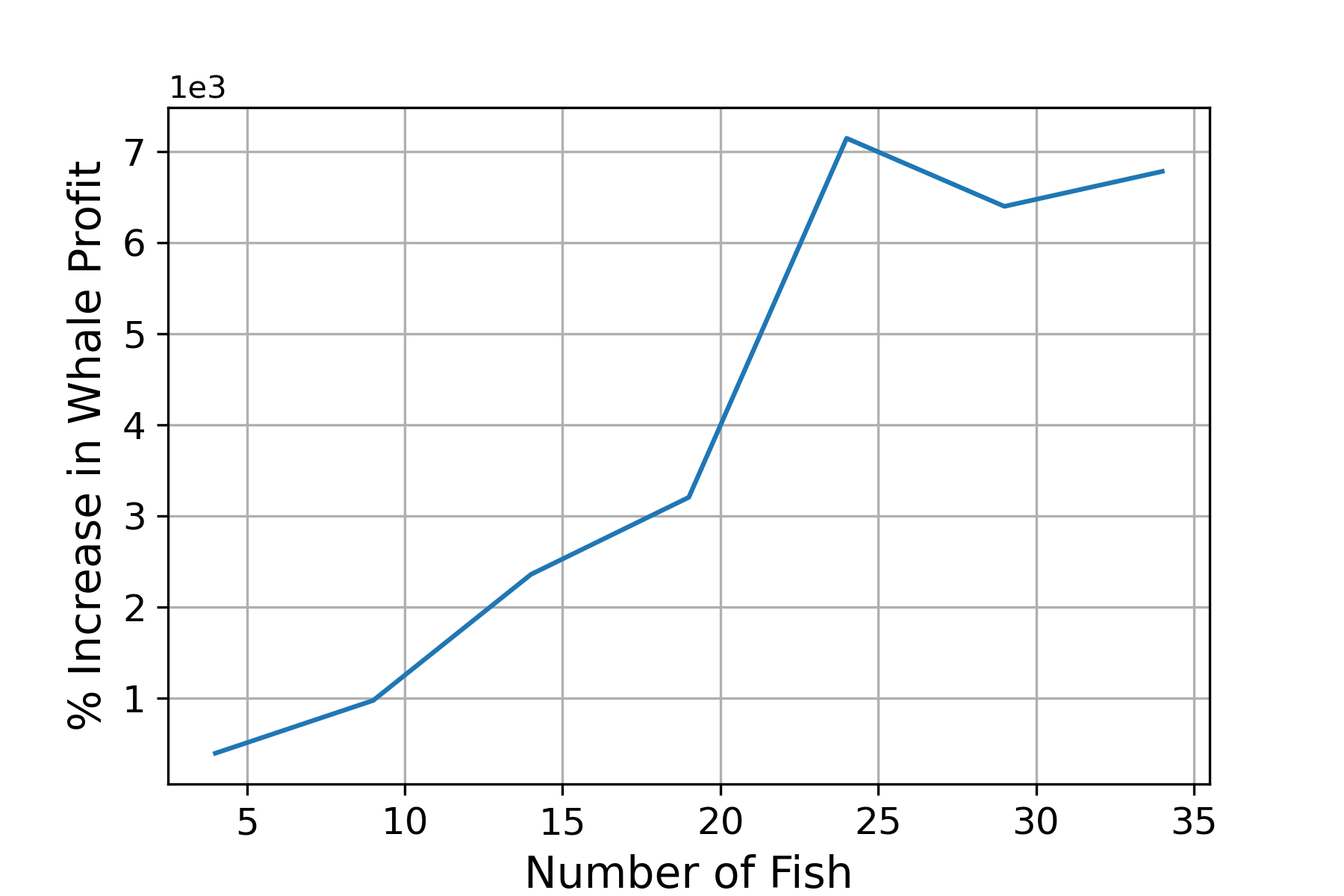}
	\caption{Percent increase in whale strategy and whale profit versus the
		number of fish when compared to the unconstrained equilibrium strategy and
		profit.}
	\label{fig:scenario3}
\end{figure}

In figure~\ref{fig:scenario3}, we consider the setting where there is one
player who has unlimited budget (whom we will call a \emph{whale}) and all
remaining players have some budget $M_i < q/n$ (these players are referred to
as \emph{fish}). The budgets of the fish are drawn uniformly from the interval
$M_i \sim [0, q/n]$ and the initial strategy of each fish is drawn uniformly at
random from the interval $[0, M_i]$. The equilibrium strategy chosen by the
fish is to use their entire budget, while the whale chooses a strategy in
excess of the unconstrained equilibrium strategy and is, as a result, able to
extract greater profit. Figure~\ref{fig:scenario3} illustrates that the whale
chooses an increasingly large strategy and receives an increasing
profit as the number of fish increases.

\begin{figure}
	\centering
	\includegraphics[width=0.45\linewidth]{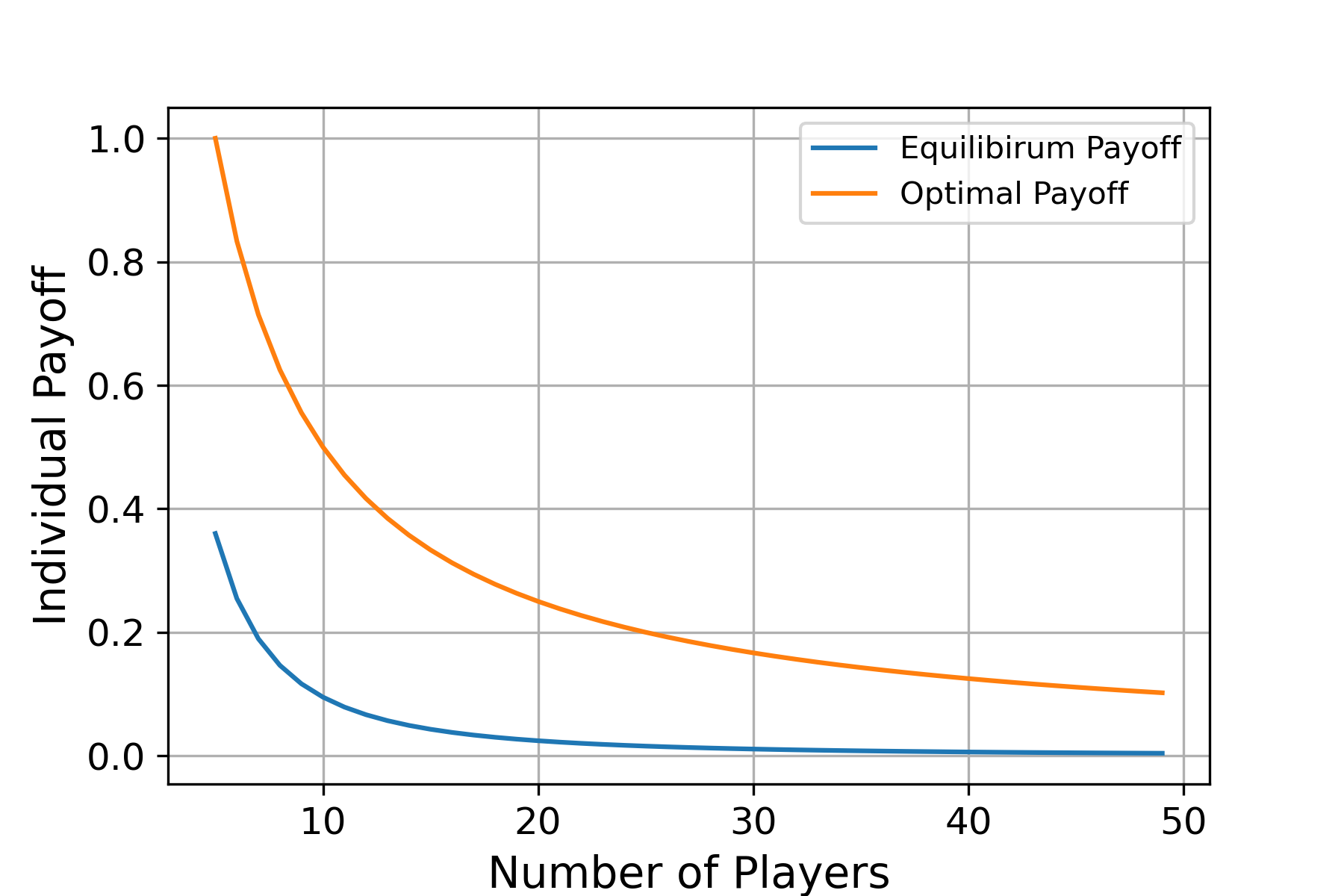}
	\includegraphics[width=0.45\linewidth]{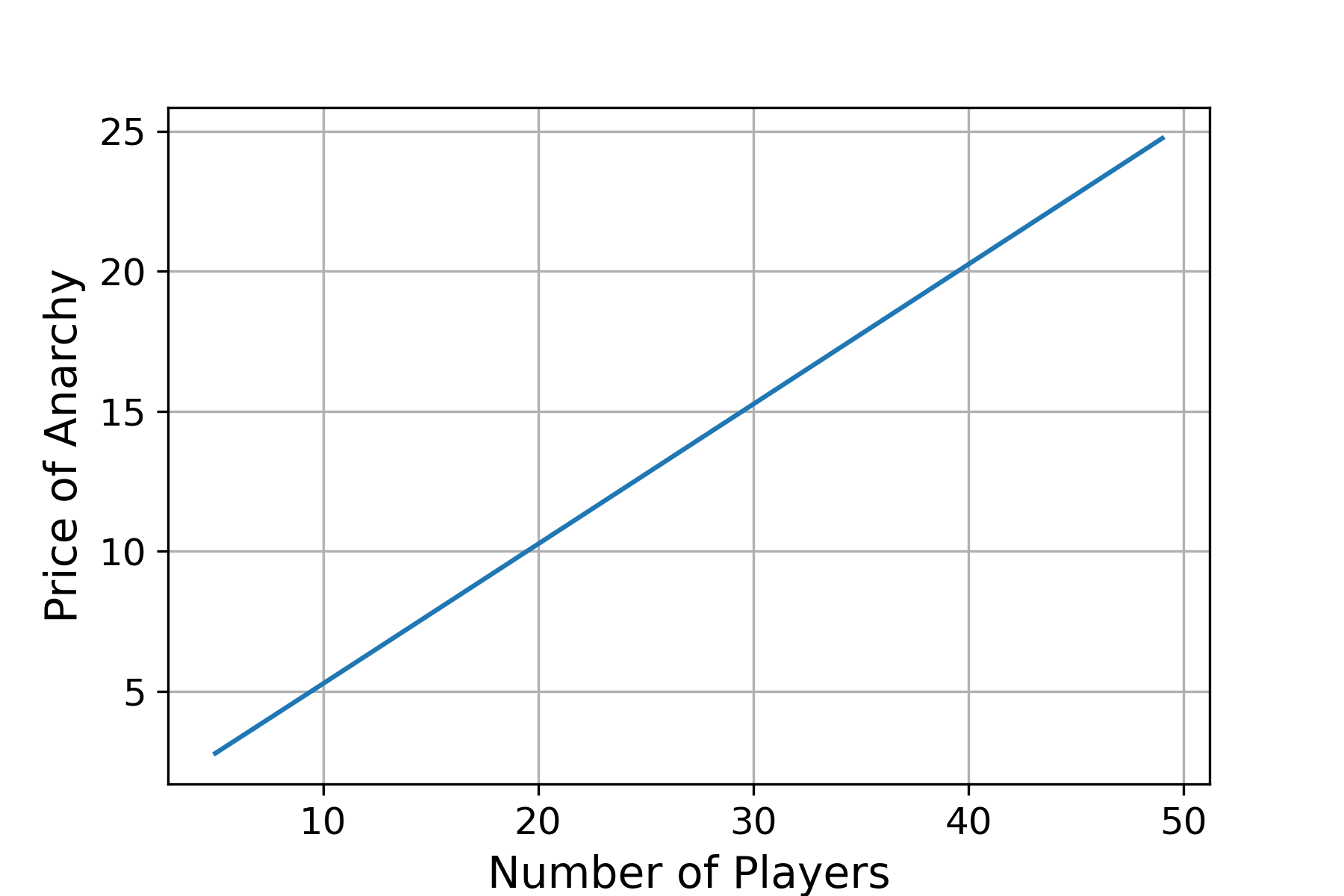}
	\caption{(Left) Individual payoff of a player versus the number
		of players. (Right) Ratio of the optimal payoff divided by the equilibrium
		payoff versus the number of players.}
	\label{fig:anarchy}
\end{figure}

\paragraph{Price of anarchy.} In~\S\ref{sec:pro-rata-game} we established
the order of growth of the price of anarchy. Here we illustrate the price of
anarchy numerically for the specific family of payoff functions introduced previously
in this section. We again fix $\gamma=0.99, R_1 = 200, R_2 = 250$ and $c=1$.
The left plot of figure~\ref{fig:anarchy} illustrates the optimal payoff function and
the equlibrium payoff function as a function of the number of players $n$ while the
right plot of figure~\ref{fig:anarchy} illustrates the price of anarchy as function of $n$.

\section{Additional Numerics} \label{app:additional_numerics}
Here we expand on the simulations introduced in appendix \ref{app:numerics} using a
class of utility function that allows us to express many quantities of interest in
closed form.

\paragraph{Game setup.}
We consider the following three scenarios:
\begin{enumerate}
	\item At iteration $t$, player $i$ takes action equal to the best response to $x_{-i}^{t-1}$.
	\item At iteration $t$, player $i$ takes action equal to the best response to $x_{-i}^{t-1}$ subject to a bounded update constraint ($\vert x_i^t - x_i^{t-1}\vert \leq \delta$).
	\item At iteration $t$, player $i$ takes action equal to the best response to $x_{-i}^{t-1}$ subject to a budget constraint ($x_i^t \in [0, M_i]$).
\end{enumerate}

\paragraph{Payoff functions.} For these simulations, we use functions $f$ of
the form $f(t) = t^\beta-\gamma t$ where $0 < \beta < 1$ and $\gamma > 0$. Note
that $f$ is concave as it is the sum of two concave functions and $f(0) = 0$.
These functions also satisfy the strict concavity property~\eqref{eq:strict}
since
\[
f(\alpha t)=\alpha^\beta t^\beta-\alpha\gamma t > \alpha t^\beta-\alpha\gamma t = \alpha f(t),
\]
for $0 < \alpha < 1$.

\paragraph{Shared equilibrium.} 

% I don't think we need these since we can write the equilibrium directly.
%The conditions provided
%in~\S\ref{sec:symmetric_equi} which guarantee a unique, finite solution for
%problem~\eqref{eq:shared-problem} hold for this particular choice of payoff
%function. To see this, note that $f'(t) > 0$ for all small enough $t > 0$,
%and $f(0) = 0$, so there is some small value $z > 0$ with $f(z) > 0$.
%If we set $w = \gamma^{1/(\beta - 1)}$, then it is easy to
%verify that $f(w) = 0$.

The (unique) symmetric pure equilibrium strategy is the solution to
problem~\eqref{eq:shared-problem}. This is easy to compute using the first
order optimality conditions for problem~\eqref{eq:shared-problem} given
in~\eqref{eq:opt-condition}. Plugging in this particular form of $f$, we have:
\[
(n-1)(q^\beta - \gamma q) + q(\beta q^{\beta - 1} - \gamma) = 0,
\]
which has a solution
\[
q = \left(\frac{\beta + n - 1}{n\gamma}\right)^{1/(1-\beta)}.
\]
The equilibrium is then given by $x_i = q/n$, for each player $i=1, \dots, n$.

\begin{figure}
	\centering
	\includegraphics[width=0.45\linewidth]{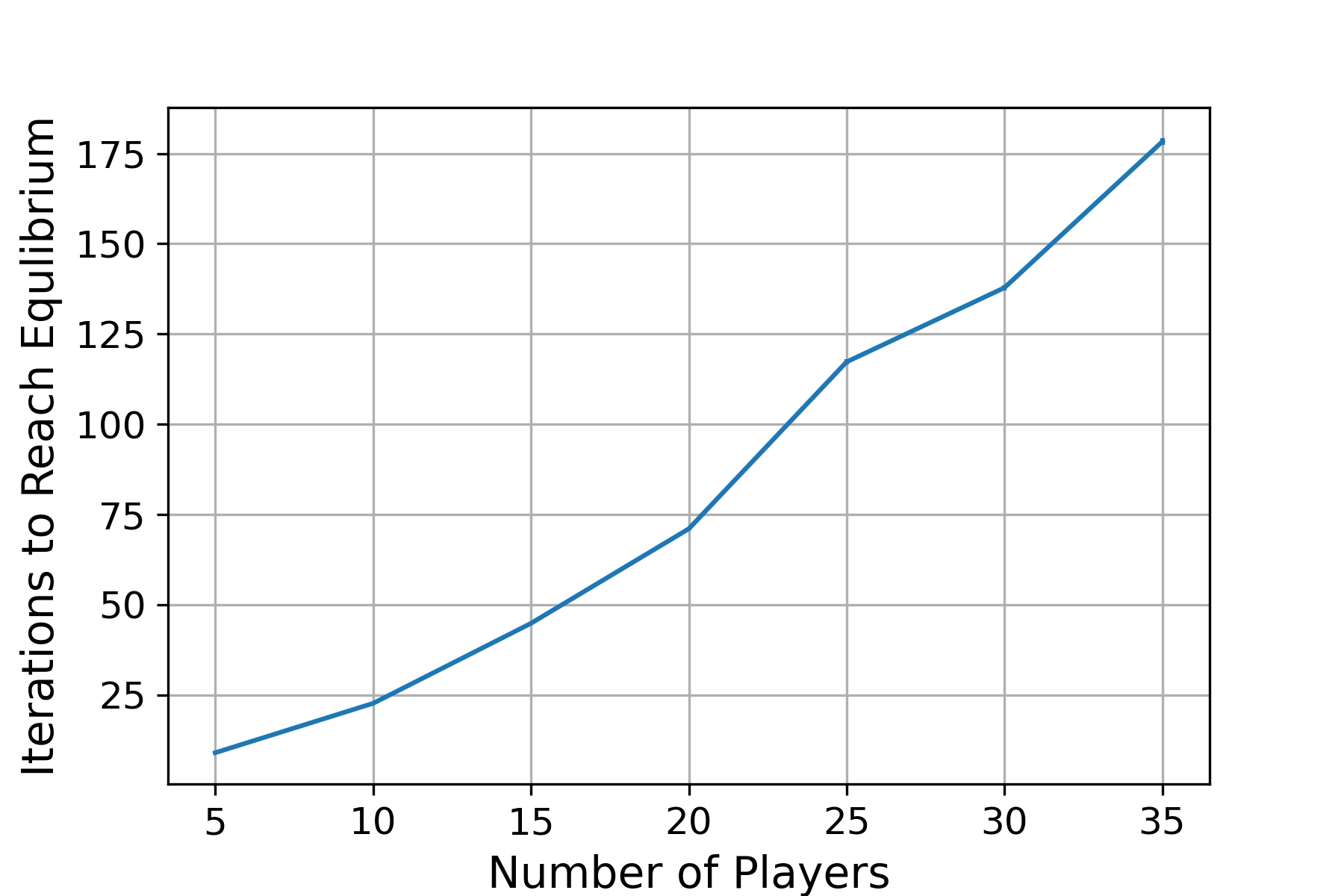}
	\includegraphics[width=0.45\linewidth]{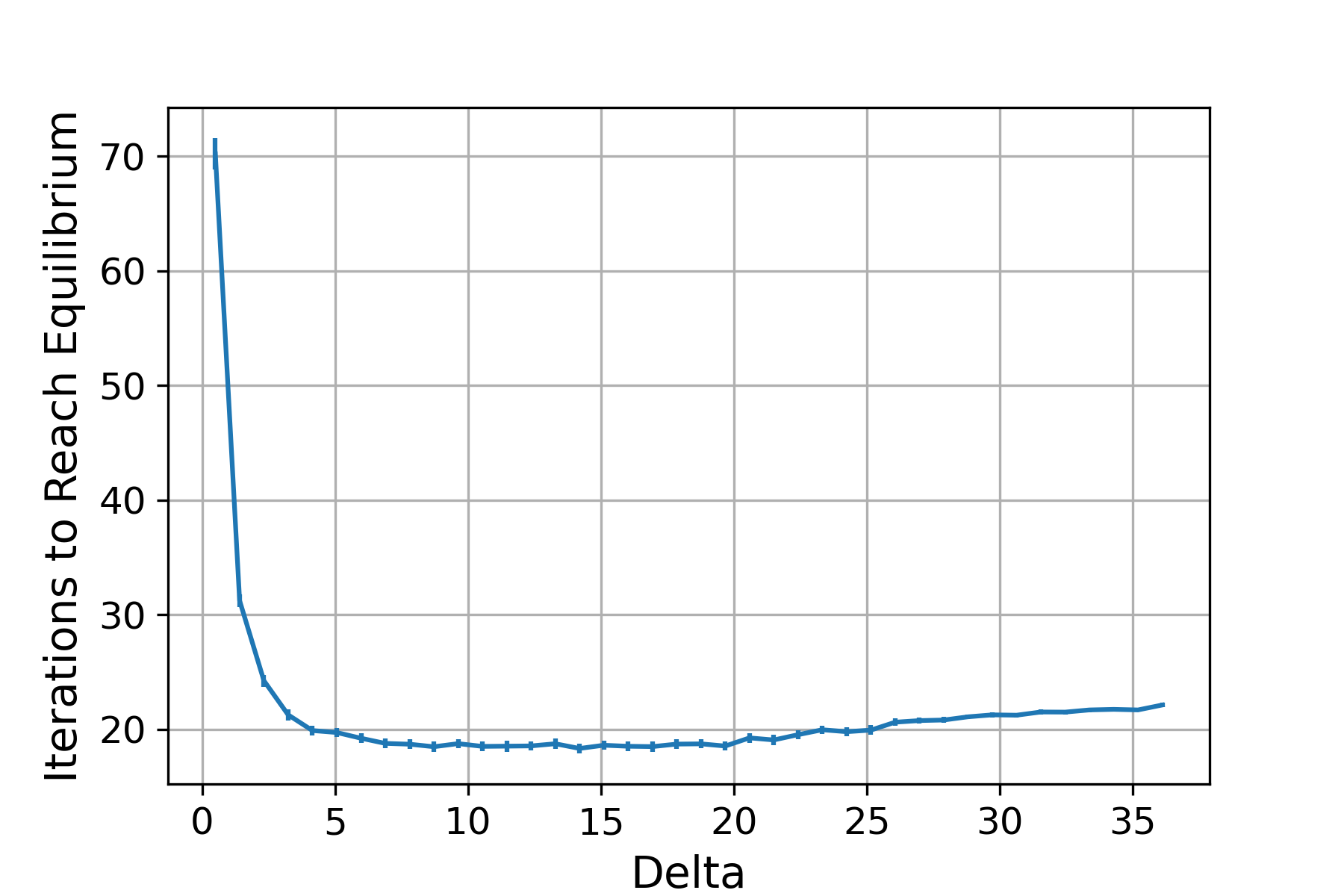}
	\caption{(Left) Number of iterations to reach equilibrium versus the number
		of players in Scenario 1. (Right) Number of iterations to reach equilibrium
		versus $\delta$ in Scenario 2 (with $n=10$ players).}
	\label{fig:scenario1+22}
\end{figure}
\paragraph{Simulation results.} In our simulations, we fix $\beta=0.5$ and
$\gamma=0.05$. We average each reported value over $100$ trials. In
figure~\ref{fig:scenario1+22}, the intial strategy of each player is drawn
uniformly at random from the interval $(0, w/n)$, where $w$ is a value such
that $f(w) = 0$.

The left plot of figure~\ref{fig:scenario1+22} illustrates that the number of
iterations needed to reach the unique equilibrium, in the absence of budget
constraints, scales superlinearly in the number of players. The right plot
demonstrates that in the scenario of bounded strategy updates, for small values
of $\delta$, the number of iterations required to reach equilibrium increases
significantly when compared to the unbounded strategy update scenario.

\begin{figure}
	\centering
	\includegraphics[width=0.45\linewidth]{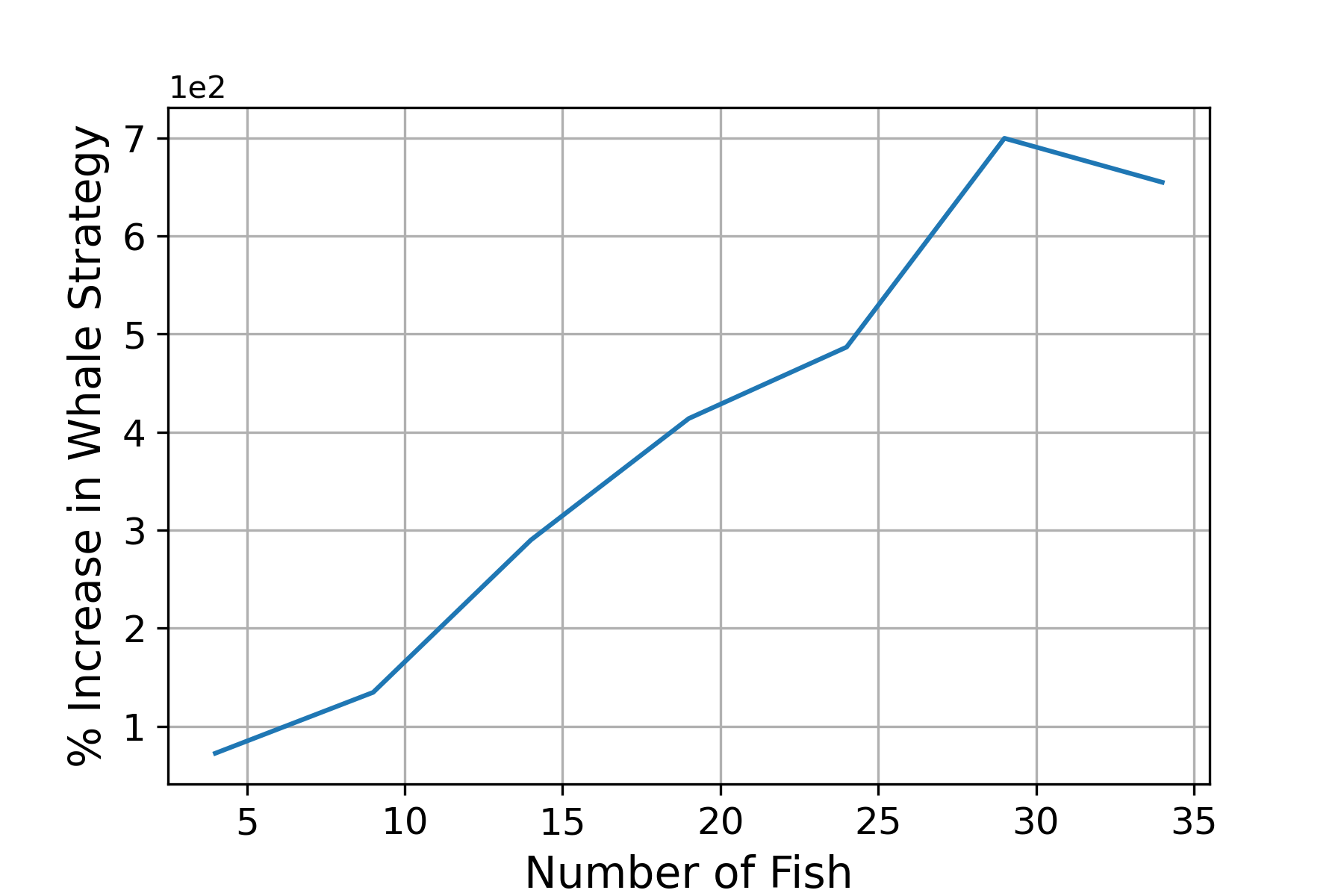}
	\includegraphics[width=0.45\linewidth]{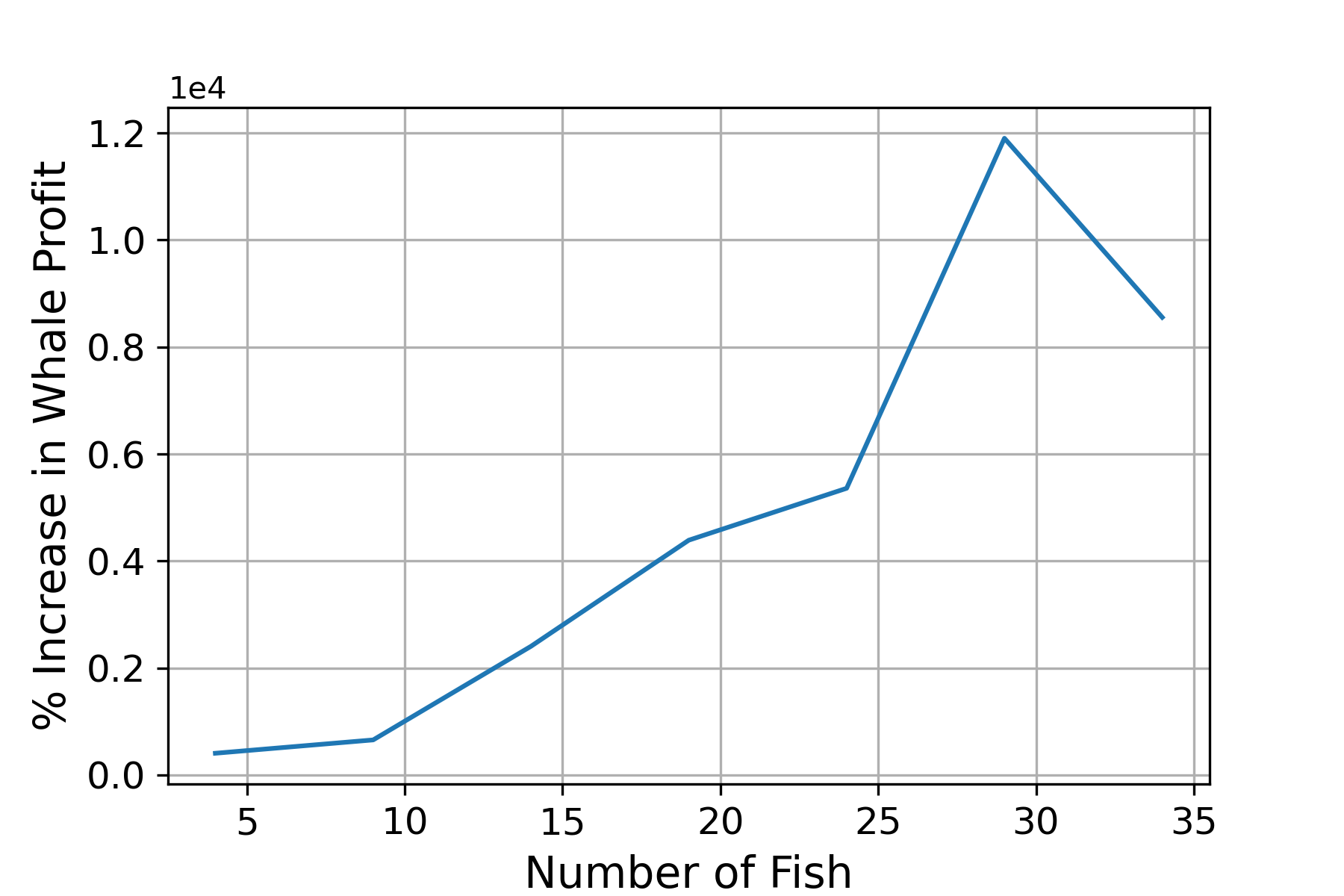}
	\caption{Percent increase in whale strategy and whale profit versus the
		number of fish when compared to the unconstrained equilibrium strategy and
		profit.}
	\label{fig:scenario33}
\end{figure}

In figure~\ref{fig:scenario33}, we consider the setting where there is one
player who has unlimited budget (whom we will call a \emph{whale}) and all
remaining players have some budget $M_i < q/n$ (these players are referred to
as \emph{fish}). The budgets of the fish are drawn uniformly from the interval
$M_i \sim [0, q/n]$ and the initial strategy of each fish is drawn uniformly at
random from the interval $[0, M_i]$. The equilibrium strategy chosen by the
fish is to use their entire budget, while the whale chooses a strategy in
excess of the unconstrained equilibrium strategy and is, as a result, able to
extract greater profit. Figure~\ref{fig:scenario33} illustrates that the whale
chooses an increasingly large strategy and receives an increasing
profit as the number of fish increases.

\paragraph{Price of Anarchy}

\begin{figure}
	\centering
	\includegraphics[width=0.45\linewidth]{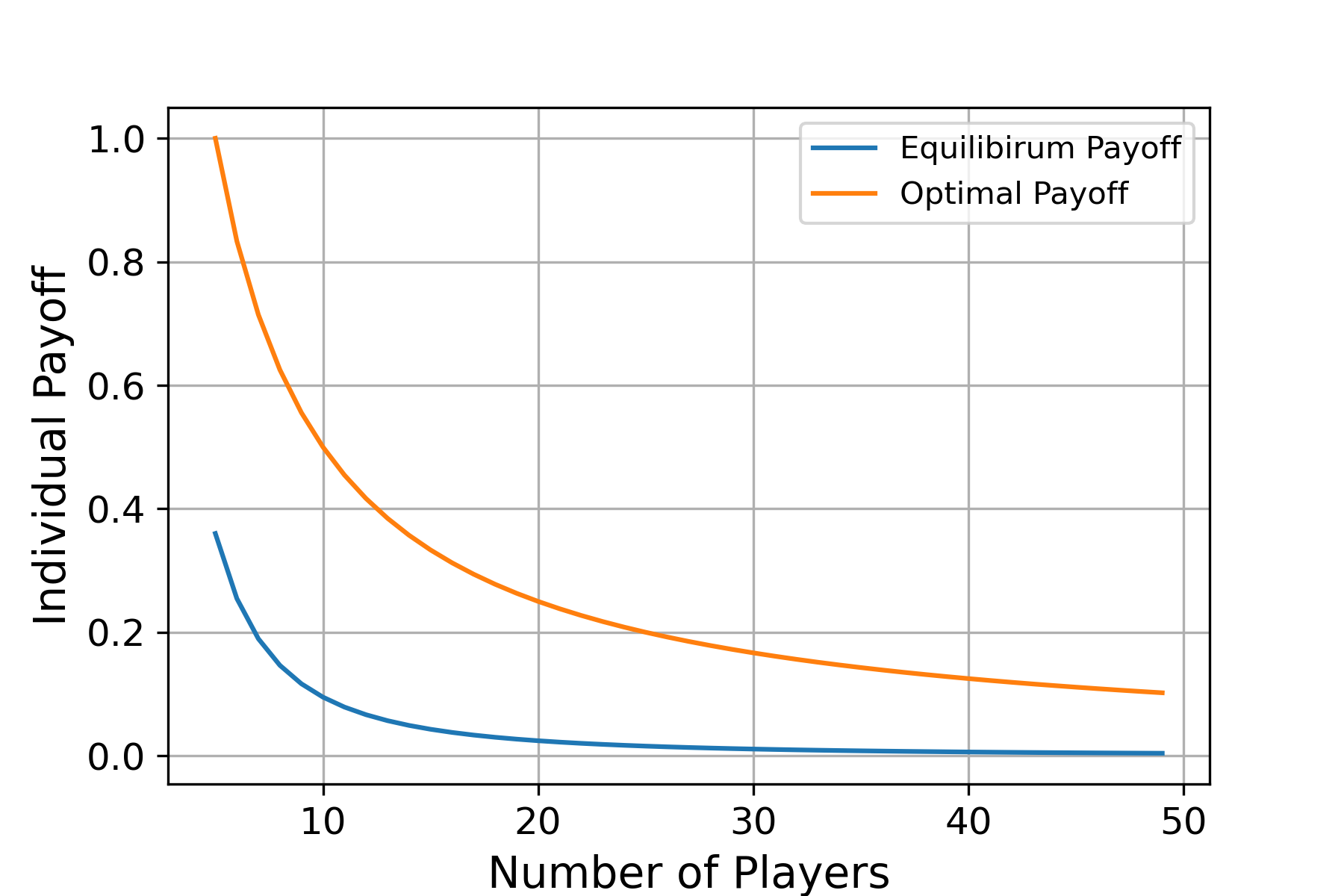}
	\includegraphics[width=0.45\linewidth]{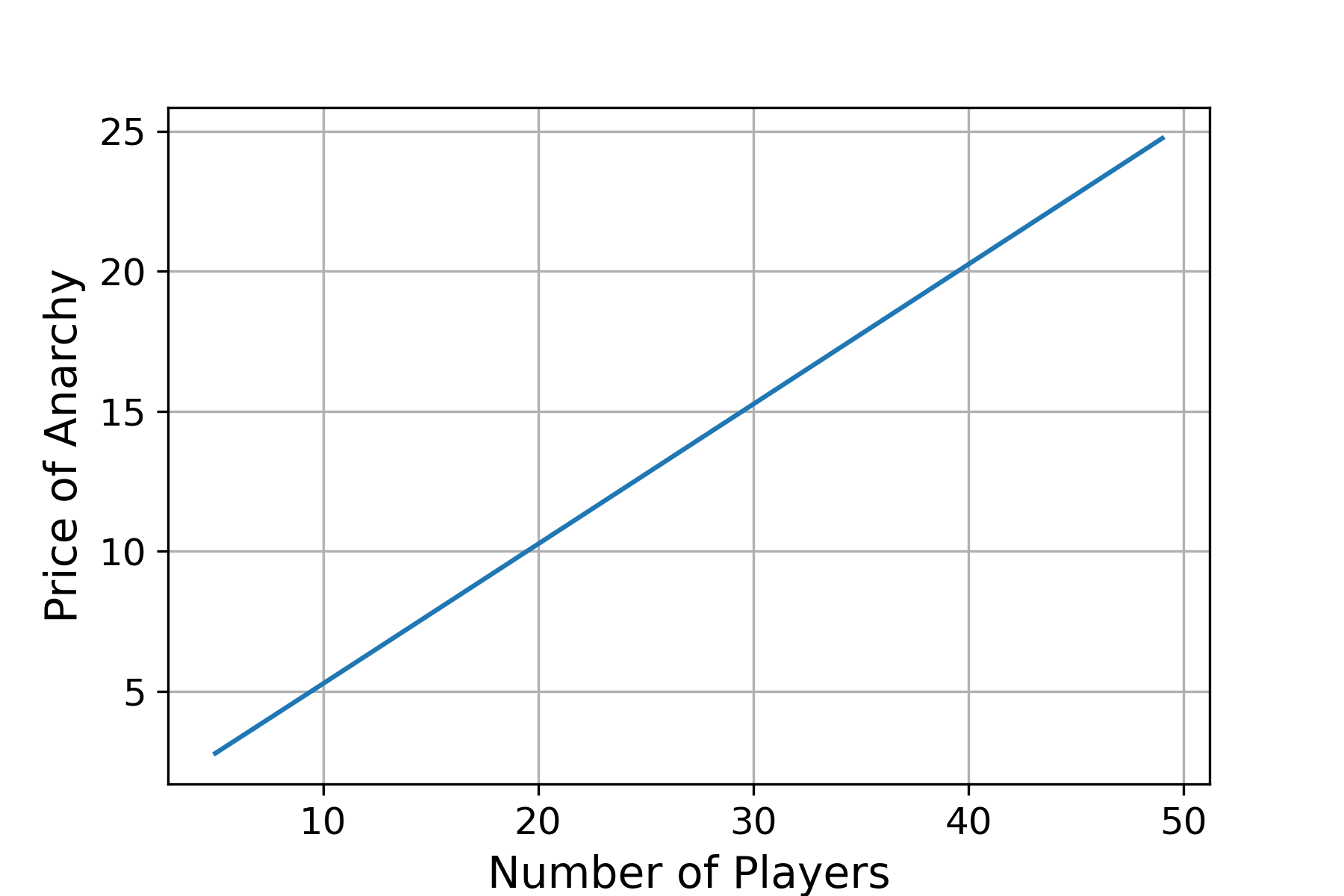}
	\caption{(Left) Individual payoff of a player versus the number
		of players. (Right) Ratio of the optimal payoff divided by the equilibrium
		payoff versus the number of players.}
	\label{fig:anarchy3}
\end{figure}

The equilibrium payoff can easibly be found to be
\[
\frac{1}{n}f(q) = \bigg{(}\frac{n+\beta-1}{\gamma n}\bigg{)}^{\beta/(1-\beta)} \bigg{(}\frac{1-\beta}{n^2} \bigg{)}.
\]
Similarly, it can be show that the optimal payoff conditioned on every agent receving the same
payoff is given by
\[
\frac{1}{n} \sup f = \bigg{(} \frac{\beta}{\gamma} \bigg{)}^{\beta/(1-\beta)} \bigg{(}\frac{1-\beta}{n} \bigg{)}.
\]
We obtain the price of anarchy by taking the ratio of the equilibrium payoff and the optimal payoff:
\[
\frac{\sup f}{f(q)} = n \bigg{(} \frac{\beta n}{n + \beta - 1} \bigg{)}^{\beta / (1-\beta)}.
\]
We again fix $\beta=0.5$ and $\gamma=0.05$. The left plot of figure~\ref{fig:anarchy3}
illustrates the optimal payoff function and the equlibrium payoff function as a function
of the number of players $n$ while the right plot of figure~\ref{fig:anarchy3} illustrates
the price of anarchy as function of $n$.

\section{Relaxing strict concavity} \label{app:strict-concavity}
We do not, in fact, need strict concavity in the proofs above.
Instead, we only need that $f$ has `some curvature' at $0$.
Specifically, it suffices that for all $t$ and $t'$ such that $0 < t < t'$,
we have
\[
f(t) > \frac{f(t')}{t'} t.
\]
Written in English, this is the condition that the chord from $0$ to $t$ always
lies strictly below the function. This condition is sometimes difficult to
confirm for general functions $f$, so we will show that this is equivalent to
the (potentially simpler-to-handle) property that all supergradients at $0$ lie
strictly above the function at all points. We will show that, for any concave
function $f: \reals_+ \to \reals$ with $f(0) = 0$, the following two statements
are equivalent: (a) there is some $s' > 0$ and $\alpha \in \reals$ such that for
every $s$ with $0 \le s \le s'$ we have
\[
f(s) = \alpha s,
\]
and (b) there exists some $0 < t < t'$ such that
\begin{equation}\label{eq:condition}
	\frac{f(t)}{t} = \frac{f(t')}{t'}.
\end{equation}
The statement above follows from the negation of both (a) and (b). This
equivalence has a simple interpretation: if the point $(0, 0)$ is collinear
with any other two points on the graph of $f$, $\{(s, f(s)) \mid s > 0\}$, then
the function $f$ is a piecewise function with a linear segment starting at $0$.
The converse of this is that if the function $f$ has no linear segment around
$0$ (\ie, every linear overestimator around $0$ lies strictly above $f$) then
any chord must lie strictly below the function.

\paragraph{Proof.}
The forward implication is very easy: pick $t' = s'$ and let $t$ be any $0 < t < s'$,
then we have
\[
\frac{f(t')}{t'} = \alpha = \frac{f(t)}{t}.
\]
Now we'll consider the reverse implication. Given $0 < t < t'$
satisfying~\eqref{eq:condition}, we will show that, for any $0 \le s \le t$ we
have
\[
f(s) = \frac{f(t)}{t}s,
\]
which satisfies the original claim with $\alpha = f(t)/t$. First, it is easy to show
that
\begin{equation}\label{eq:cond-inequality}
	f(s) \ge \frac{f(t)}{t}s,
\end{equation}
since
\[
f(s) = f\left(\frac{s}{t}t + \left(1-\frac{s}{t}\right)0\right) \ge \frac{s}{t}f(t),
\]
where the inequality follows from the concavity of $f$ and the fact that $f(0) = 0$.
We will now show that any function $f$ satisfying~\eqref{eq:cond-inequality} strictly, \ie,
\begin{equation}\label{eq:contrapositive}
	f(s) > \frac{f(t)}{t}s,
\end{equation}
for some $0 < s < t$ cannot be concave. The result follows from the
contrapositive. To see this, let $0 < \gamma \le 1$  such that $t = \gamma s +
(1-\gamma)t'$, then
\[
\gamma f(s) + (1-\gamma)f(t') > \gamma\frac{f(t)}{t}s + (1-\gamma)\frac{f(t)}{t}t' = f(t)
= f(\gamma s + (1-\gamma)t'),
\]
so $f$ cannot be concave. The inequality follows directly from
conditions~\eqref{eq:condition} and~\eqref{eq:contrapositive}, and both the first
and second equalities follow from the definition of $\gamma$.

\section{Rosen condition}\label{app:rosen}
Pro-rata games, even concave ones, do not satisfy the Rosen
condition~\cite{rosen1965} for the uniqueness of equilibria in concave games.
The Rosen condition for uniqueness is that if, there exists some $z \ge 0$ with $z
\ne 0$ such that
\[
\Phi(x) = \begin{bmatrix}
	z_1 \partial_1 U_1(x)\\
	\vdots\\
	z_n \partial_n U_n(x)
\end{bmatrix}
\]
is a strictly monotone operator; \ie, for any $x \ne y$ we have
\[
(y - x)^T(\Phi(y) - \Phi(x)) > 0,
\]
then there is a unique equilibrium that is also pure. (Here, $\partial_i$
denotes the $i$th partial derivative.) This is a common condition used to prove
the uniqueness of pure equilibria in games. We will show that this condition
does not hold in general for concave pro-rata games, even under most `niceness'
assumptions such as strict concavity or even strong concavity and
differentiability.

Setting $2x = y = \ones$ then the condition can be written as (using the
definition of $U$)
\[
(\ones^Tz/2)((1/n)(f'(n) - f'(n/2)) + (1-1/n)(f(n) - 2f(n/2))) > 0,
\]
but this can be rewritten (since $\ones^Tz > 0$)
\[
(1/n)(f'(n) - f'(n/2)) + (1-1/n)(f(n) - 2f(n/2)) > 0,
\]
which is clearly not true for all concave functions $f$, since picking $f(t) =
\min\{t, 3n\}$ suffices. (A mollifying argument would show that this also gives
a reasonable counterexample even in the case that $f$ is strictly concave and
differentiable.) A more direct counterexample that is differentiable and
strictly concave is $f(t) = (4n)^2 - (4n - t)^2$, which is slightly more
difficult to verify.

\end{document}